# Study of silicon photomultipliers for use in neutron decay experiments

D. Dubbers

*Physikalisches Institut, Heidelberg University, Im Neuenheimer Feld 226, 69120 Heidelberg, Germany*

ABSTRACT

Photon readout of plastic scintillators is investigated with the aim of improving the precision of neutron $\beta$ decay experiments. Neutron decay is nowadays studied with high statistics, based on up to $10^9$ registered decay events, and leads to strongly improved limits on new physics beyond the standard model, with $\beta$s often registered in plastic scintillators. The main systematic errors in these experiments are due to imperfect characterization of the scintillators with respect to linearity, energy resolution, and electron backscattering. We study whether these errors can be diminished when the conventional photomultipliers used for scintillator readout are replaced by silicon photomultipliers (SiPMs). To this end, various theoretical and experimental tools are developed, and a procedure for handling the extreme dark rates of SiPMs is proposed. In $\beta$ spectroscopy, so the conclusion, plastic scintillator readout with SiPMs can significantly improve energy response, and help providing reliable corrections for electron backscattering.

## 1. Introduction

In the past few years, the precision of data on neutron $\beta$ decay has considerably improved. Limits on deviations from the vector minus axial-vector (V–A) structure of the electroweak standard model (SM) derived from $\beta$ decays have evolved within a decade from order 10% [1] to below $10^{-3}$, see [2] and references therein. These limits are now better than the corresponding limits from high-energy searches [3, 4]. To improve limits even further, one must reduce the remaining systematic errors, which are mainly due to insufficient detector characterization.

The majority of neutron decay experiments use plastic scintillators for the spectroscopy of the decay electrons. Recent examples are the instruments PERKEO at ILL [5], UCNA at LANL [6], aCORN [7] and emit [8] at NIST, nTRV [9] at PSI, the upcoming PERC at FRM II [10], and the projected BRAND [11]. Neutron decay volumes have sizes of up to several liters, while the scintillator plates have lateral dimensions of up to several decimeters, with typical thickness of 5 mm. Magnetic guiding fields of order Tesla transport the electrons from the decay volume to their detectors.

The advantages of plastic scintillators are their thin dead layer, fast response time, small background sensitivity, and low electron backscattering. Their main disadvantage is a mediocre energy resolution, mainly due to the statistical fluctuations of the number of photons detected from the scintillator. The photons usually leave the scintillator plates laterally through their narrow side faces, and are transported to some distant photomultiplier pubes (PMTs) via acrylic light guides.

In the precision experiments listed above, the dominant sources of systematic error are insufficiently known energy response, as well as electron backscattering from the scintillators [12]. These errors are also predominant in searches for new physics beyond the SM, as in the experiments on a non-SM tensor coupling via a Fierz interference term [13, 14, 15]. The aim of the present article is to find out whether these errors can be diminished by replacing the PMTs for photon readout by silicon photomultipliers (SiPM) attached directly to a scintillator's narrow side faces, with no need for light guides.

SiPMs are solid-state photodetectors of typically 3×3 mm$^2$ area, composed of about $10^4$ single photon avalanche diodes of area 25×25 μm$^2$, each such pixel acting as a Geiger-Müller counter. In $\beta$ spectroscopy with scintillators, there is no need for spatial resolution, and light intensity is low enough that the photoelectron signals from the individual pixels in the SiPM can simply all be added up, thus directly counting the number of photons with high linearity. In neutron decay, count rates are far from saturating the SiPMs, in which case the intrinsic linearity of SiPMs is close to ideal.

SiPMs are superior to PMTs with respect to their small size, weight, and cost, their low operating voltage, their insensitivity to magnetic fields, and their linearity. They are at least equivalent to PMTs concerning signal rise time and single photon detection. However, the typical dark rates of SiPMs of up to 100 kHz per mm$^2$ detector area look rather prohibitive, as compared to order 100 Hz per cm$^2$ for PMTs.

In recent years, SiPMs have found many applications, in particular in high energy-, astro-, and medical physics. For reviews on properties and characterization of SiPMs, see [16, 17, 18, 19]. Their use in nuclear and particle physics is reviewed in [20], and applications in high-precision γ spectroscopy are described in [21], [22].

The paper is organized as follows. In Section 2 various photon loss mechanisms in scintillators are investigated with a simple method of virtual images to model photon propagation. Section 3 describes the experimental setup and Section 4 the measurement of the photoelectron yield of the scintillator-plus-SiPM system. Section 5 shows how distortions of line spectra due to electron backscattering from their detector can be corrected for. To reduce the number of free parameters in our studies, photon absorption and reflection losses in the scintillators used are reported in Section 6. In Section 7, for two scintillator sizes the photoelectron yield is measured as a function of the number of SiPMs installed, and is compared to expectations to enable extrapolation to larger scintillators as required in neutron decay experiments. In Section 8 a strategy for the suppression of the extreme dark rates expected for the larger scintillator setups is proposed. Our first application of SiPMs will be in a time-of flight experiment on electron backscattering, described in Section 9. A summary follow in Sections 10.

## 2. Modeling of the combined scintillator-SiPM system

Photons generated in a plastic scintillator are repeatedly reflected from the scintillator's inner surfaces, or from additional reflector foils, until they are absorbed in a detector attached to the scintillator. Seemingly, an ideal gas escaping from a vessel through a leak resembles this process, but the well-known formulae for this case do not apply because the photons in the scintillator are not an ergodic system in thermal equilibrium.

**Fig. 1a** shows a thin scintillator plate (or "tile") with an SiPM attached to one of its narrow outer faces. This is the configuration used in all our studies, although our model works for any configuration of a scintillator in the shape of a box (i.e., an arbitrary rectangular parallelepiped) coupled to a photon detector of rectangular shape. For a critical angle $\theta_c$ near 45°, total reflections of scintillator light almost never occur at these narrow faces [23]. Therefore, these faces must be covered by reflecting foils, with openings for the SiPMs. In contrast, the large uncovered faces through which the electrons enter are totally reflecting from the inside only for angle $\theta > \theta_c$, the small reflection probability for $\theta < \theta_c$ being also neglected (with $\theta$ measured against the normal of the face).

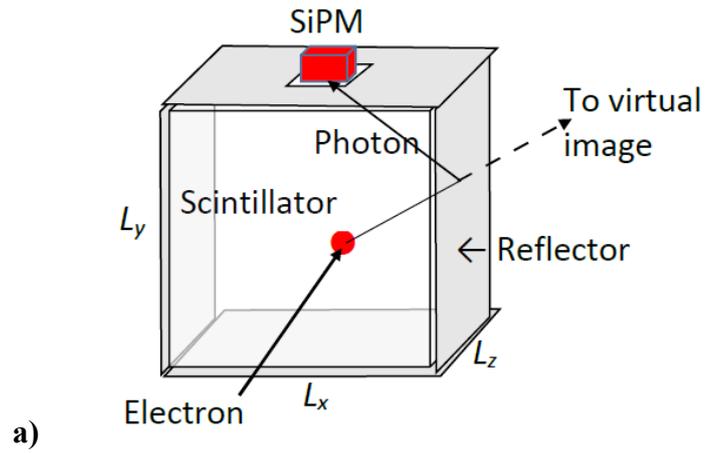

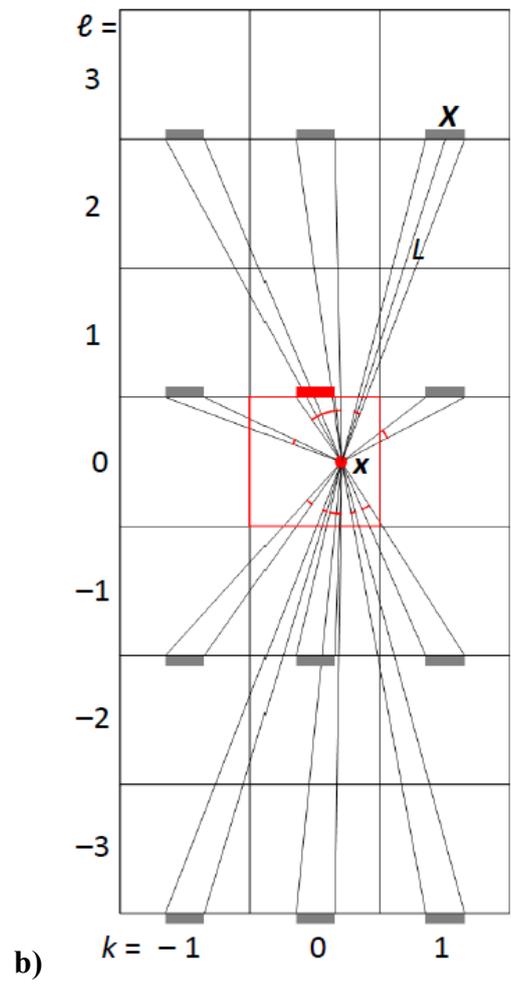

**Fig. 1.** Photon trajectories in a scintillator-plus-detector system. **a)** Scintillator plate with reflectors (gray) and a photon detector (red) attached to one of its narrow side faces. **b)** The same system treated by the method of virtual images. The real system in the center (red) is surrounded by its virtual images (gray), as mirrored by the reflectors. The small arcs surrounding the photon emission point (both red) indicate the angular ranges under which the photons that survive losses can be detected in a given detector or its virtual image.

There are many publications on ray tracing Monte Carlo simulations of scintillators, see [24] and references therein. For the frequent case of scintillators in the shape of a rectangular plate, photon trajectories were also investigated algebraically, see [23] and [25]. In the present paper, a simple method of virtual images is used instead. For two dimensions, the method is visualized in **Fig. 1b**. In the center is the scintillator plate in red, surrounded by an (in principle infinite) set of virtual images in gray as seen through the mirroring narrow side faces. The scintillator tiles are numbered along $x$, $y$, $z$ by integers $k$, $\ell$, $m$, with the central real tile at $k = \ell = m = 0$. Fig. 1b shows the $x$-$y$ plane for $z = 0$.

The detection probability for a photon, generated at position $x$ within the scintillator, depends on the sum of solid angles under which the real and virtual detectors appear when seen from position $x$. In the figure these angles are indicated by the circular segments drawn around $x$. Note that a detector, real or virtual, frequently shadows some of the more distant virtual detectors. With this model, the problem of light evolution in a plate is reduced to a simple addition of angles, with photon losses taken into account as follows.

A photon emitted in the direction of a virtual SiPM at position $X$, attached to a tile numbered $(k,\ell,m)$, will cross several boundary faces between real or virtual adjacent tiles. If the boundary is between two narrow faces of the scintillator, as in Fig. 1b, this corresponds to a reflection on a reflector foil of reflectivity $R_f \leq 1$. The number of these crossings equals $|k|+|\ell|$, and leads to a mean photon loss $1 - (R_f)^{|k|+|\ell|}$. If the boundary is between two large faces of the scintillator, this corresponds to a total reflection on an inner surfaces of the scintillator, of reflectivity $R_t \leq 1$ for angle of incidence $\theta > \theta_c$, and $R_t = 0$ for $\theta \leq \theta_c$. The number of these crossings equals $m$, and leads to a mean photon loss $1 - (R_t)^m$. For given $k$ and $\ell$, the number $m$ of total reflections is limited by $\theta_c$ to $m_{max} = \text{floor}[L_0/(L_z \tan\theta_c)]$, where floor indicates rounding to the greatest integer less than the argument, with $L_0 = \sqrt{(kL_x)^2 + (\ell L_y)^2}$ for scintillator dimensions $L_x$, $L_y$, $L_z$ as in Fig. 1a. In addition there are absorption losses of mean size $1 - \exp(-L/\lambda)$, where $L = |X - x|$ is the length of the photon trajectory and $\lambda$ the photon absorption length of the scintillator.

In the following we denote the photoelectron yield (p.e./MeV) shortly as yield $R$. In **Figs. 2** and **3**, $R$ is calculated as a function of various parameters. With the refractive index of the plastic scintillator $n = 1.57$, as used in the experiments described below, the critical angle for total reflection is $\theta_c = \arcsin(1/n) = 40°$. The scintillator's transmission parameters are chosen to $R_f = R_t = 0.95$ and $\lambda = 100$ cm, unless stated otherwise, and the area of photon detectors to 3×3 mm² each. The calculations show that, for the setup shown in Fig. 1, $R$ depends only little (within several percent) on photon start position $x$ or detector position on the narrow faces, which therefore are chosen to $x = 0$ and $X = 0$, respectively.

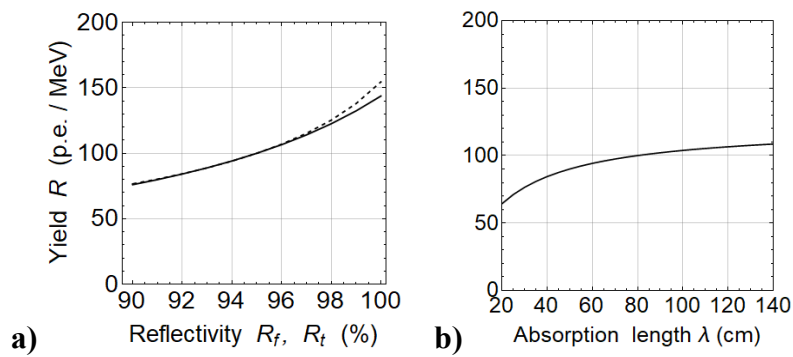

**Fig. 2.** Sensitivity of photoelectron yield $R$ to various photon loss mechanisms, for a single scintillator of dimensions 20×20×5 mm³ in a setup as in Fig. 1a. **a)** $R$ as a function of reflectivity $R_f$ of the reflector foils (solid line) and of reflectivity $R_t$ due to total reflection from the inner scintillator faces (dashed line). **b)** $R$ as a function of photon absorption length $\lambda$. In order to become independent of absolute detection efficiency, yield $R$ is arbitrarily set to 100 p.e./MeV in the center of each plot.

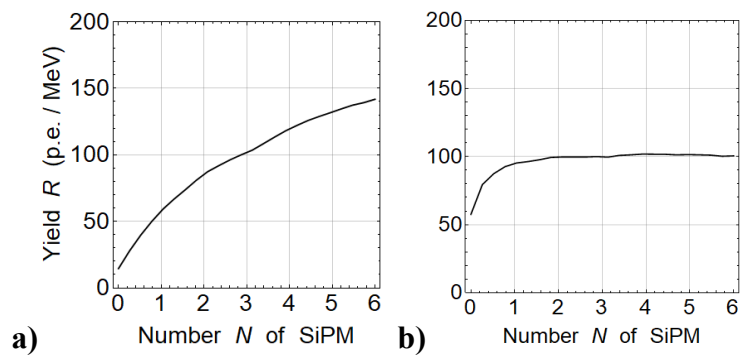

**Fig. 3. a)** Yield $R$ in dependence of the number $N$ of photon detectors installed, each of size 3×3 mm², for transmission parameters $R_f = R_t = 0.95$, $\lambda = 100$ cm, for the same geometry as Fig. 2. **b)** Without photon losses, $R_f = R_t = \lambda = 0$, yield $R$ must be independent of $N$, which here is realized for $N \gtrsim 1$. $R$ is arbitrarily set to 100 p.e./MeV in the center of each plot. For non-integer numbers $N$ of SiPMs, $R$ is calculated by varying their $x$-, but not their $y$-dimension, while real SiPMs come in the shape of squares of a given size.

**Fig. 2** shows yield $R$ in dependence of the photon transmission parameters $R_f$, $R_t$, $\lambda$. While foil reflectivity $R_f$ usually differs from total reflectivity $R_t$, their effect on yield $R$ is very similar, see Fig. 2a. The absorption length, shown in Fig. 2b, is usually large compared to the dimensions of the scintillator. **Fig. 3** shows $R$ in dependence of the number $N$ of photon detectors, installed one next to the other, in Fig. 3a with transmission parameters as stated, in Fig. 3b without transmission losses.

The number of virtual images needed in a calculation of $R$ can be obtained by setting all photon losses to zero. In this case, even a single small detector will finally capture all photons with $\theta \geq \theta_c$, with a solid angle of photon acceptance of $\Omega/4\pi = (1 - 1/n^2)^{1/2} = 0.77$. The number of virtual images must then be increased until the yield $R(N)$ becomes sufficiently independent of $N$. Fig. 3b shows $R(N)$ for the loss-free case, with $k$ and $\ell$ running up to $k_{max} = \ell_{max} \approx 40$, for which independence of $N$ is approximately reached for $N \gtrsim 1$.

Large numbers of plastic scintillators of ~centimeter size with photon readout by single SiPMs are being planned for various high-energy experiments, like in the mu3e project at PSI with about 6000 tiles in its first phase [26, 27], in the PANDA project at GSI with 10 000 tiles [28], and in the CALICE project at CERN with up to 20 000 tiles [29].

## 3. Measurement tools

For the measurement of photoelectron yield $R$ with SiPMs, the following equipment was used. The plastic scintillators, with 423 nm wavelength of maximum photon emission (BC400 from Saint-Gobain), were square tiles of two sizes, $L_x = L_y = 20$ and 40 mm, and thickness $L_z = 5$ mm. They were coupled to blue-sensitive SiPMs, with peak sensitivity at 430 nm (KETEK PM3325-WB), and with dimensions as given in the Introduction.

The SiPM chips were not glued to the plastic scintillator, so they could be reused. Instead they were pressed to the tiles with a mild force of several Newtons. The SiPMs are rather robust but can break under a stronger force, although they sometimes survive even breakage. They also forgive wrong polarity of the voltage applied, but excessive overvoltage must be avoided. Their optical contact was assured by a thin layer of silicon grease of refractive index $n = 1.41$ (Rhodosil PÂTE 7), while the glass of the SiPM entrance window has $n = 1.52$. Without grease, yield $R$ decreased by 30%. The narrow side faces of the scintillator tiles were covered with reflector foils of up to nominal 99% reflectivity (Vikuti Enhanced Specular Reflector ESR from 3m-Optical Systems). Square holes of 3.3 mm lateral dimension were punched into the foils at the positions of the SiPMs, see Fig. 1a. When aluminized mylar foils were used instead of these foils, the yield $R$ decreased by 17%.

The electronics used is simple: a detector voltage, usually 2.5 to 5 V above the breakdown voltage of 24.5 V of the diodes, was applied to the SiPMs through a serial resistor of 10 kΩ, with a parallel capacitance of 100 nF to ground. The signals were read out over a load resistor of 3 kΩ and were decoupled from this circuit by a capacitance of 100 nF. These values can be varied over a wide range without affecting the signals, because their rise time of 10 ns and decay time of 90 ns (each 10% to 90% of maximum amplitude) are due to the internal structure of the SiPM.

The typical signal amplitude of a single SiPM at overvoltage 5 V, coupled to the 20×20×5 mm$^3$ tile, is 50 mV for 1 MeV electron energy. When several SiPMs are installed in parallel, detector capacity and with it the decay time of the signal increases, while its amplitude changes only marginally. These signals then passed a timing filter amplifier (Ortec TFA 474) with integration constant set to 200 ns, with no need for a preamplifier. The spectra were registered in a multichannel analyzer (Pitaya STEMlab 125-14).

A nearly point-like $^{207}$Bi source was used, with K-conversion electron energies of $E_K$ = 0.48 MeV and 0.98 MeV, whose intensity ratio is 1 to 5.9. The source was dried onto an about 5 μm thick foil of aluminized Mylar. The electrons reached the scintillator from a distance of 12 mm through an orifice of 12 mm diameter, made from a 1.7 mm thick sheet of copper, covered towards the scintillator with 0.4 mm cardboard to absorb secondary electrons released by unconverted γ rays.

## 4. Measurement of photoelectron yield with SiPMs

For simplicity we assume that, in a line contributing to a conversion electron peak, the number $p$ of detected photoelectrons is large enough that the line spectrum can be approximated by a Gaussian,

$$\frac{\mathrm{d}p}{\mathrm{d}E} = \frac{n_0}{\sqrt{2\pi}\sigma} \exp\left[\frac{(E-E_0)^2}{2\sigma^2}\right], \qquad (1)$$

with peak position $E_0$ and variance $\sigma^2$.

In principle the mean number $\bar{p}$ of photoelectrons in a peak can be derived experimentally by dividing the measured spectral position $E_0$ of this peak by the position $E_{1pe}$ of the single photoelectron peak. In practice it is more precise to derive $\bar{p}$ from the line's position $E_0 \propto \bar{p}$ and width $\sigma \propto \sqrt{\bar{p}}$, which give $\bar{p} = E_0^2/\sigma^2$. The photoelectron yield then is $R = \bar{p}/\tilde{E}_0$ p.e./MeV, with energy $\tilde{E}_0$ given in units of MeV. The dark signals of a SiPM cannot be distinguished from the true one-photon signals, whose energy in both cases is $E_{1pe} = R^{-1}$ (for example, $R$ = 200 p.e./MeV gives $E_{1pe}$ = 5 keV). If there are not too many random coincidences, the dark counts can be suppressed by a suitable discriminator threshold.

In contrast to the above mentioned neutron decay experiments, our setup had no magnetic field to guide the electrons from the source to the scintillator. The γ rays from the $^{207}$Bi source therefore irradiated the scintillator area under the same solid angle as the conversion electrons, producing a background of 40% of the total rate. This background was measured separately, both before and after data taking, by absorbing the electrons in a sheet of 5 mm polyethylene, and was subtracted from the spectrum.

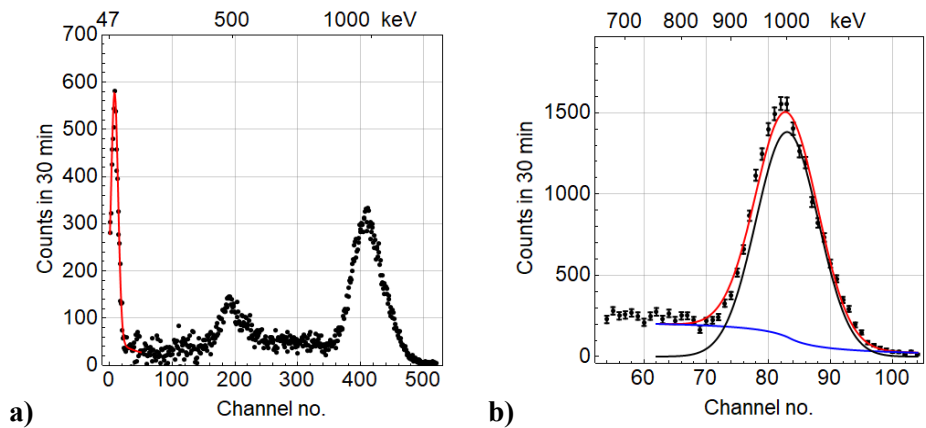

**Fig. 4.** Electron spectrum from the $^{207}$Bi source, taken with two SiPMs on a 20×20×5 mm$^3$ scintillator. The upper horizontal axes show the electron energy. **a)** The conversion electron peaks at 0.5 and 1 MeV, and the Auger electron peak at 63 keV fitted with a Gaussian Eq. (1) (red). **b)** Simultaneous fit (red) of Gaussians (black) plus backscatter Eq. (2) (blue) to the 1 MeV data (with five channels of the left figure contracted to one). From the fitted width, the yield $R$ is obtained, as described in the text.

**Fig. 4** shows a background reduced $^{207}$Bi electron spectrum, taken with two SiPMs installed on two opposite narrow faces of the 20×20×5 mm$^3$ scintillator, with an overvoltage of 5.5 V applied to the SiPM. Fig. 4a shows the total spectrum, composed of the two conversion peaks at 0.5 and 1 MeV and one Auger K-electron peak at 63 keV. With PERKEO using PMT readout, this Auger line is not fully resolved see Fig. 1 of [13]. The positions of the three peaks are linear with a zero offset of 13 channels or –47 keV, with a reduced chi-square of $\chi_0^2 = 0.92$ for two degrees of freedom.

Fig. 4b shows in red a fit to the K, L, M transitions of the 1 MeV line, based on published electron conversion data, including corrections to electron backscattering as described in the next section. The 11.8(3)% full width at half maximum (FWHM = $2\sqrt{2\ln 2}\,\sigma$) of the underlying Gaussians gives $R = 281(7)$ p.e./MeV, where the one sigma errors in the last digit, derived from the fit, are given in parentheses. Earlier PERKINO measurements with full scale PMT readout [30], in which a round scintillator of 4 cm diameter was read out through its large face via a 50 cm long light guide by a PMT of 2" diameter or 18 cm$^2$ area, had the same yield, $R \approx 300$ p.e./MeV, as the two SiPMs of 0.18 cm$^2$ area.

## 5. Correction of conversion electron line shape for backscatter effects

In Fig. 4a, some residual intensity remains between the two $^{207}$Bi conversion peaks, essentially due to electron backscattering from the scintillator. The yield $R$ obtained from the fit then depends on how these counts are continued under the measured peaks: the more residual intensity is subtracted, the smaller the width of the remaining conversion peak. As we want to know the width of the peak, and with it the yield $R$, for the undisturbed electrons, the backscatter fraction under the peak must be known. Unfortunately, its shape under the peak is not known and cannot be determined numerically. The reason for this is that the tiny energy shifts lie all within the width of the conversion peak, and backscattering near the glancing angle must be included. This involves parameter regions where simulations could be trusted only if some day experimental checks become available (hopefully, our future backscatter experiment presented in Section 9, will provide the data needed for this).

This problem can be solved empirically. In the region of interest near the measured 1 MeV peak in Fig. 4, the residual intensity $f(E)$ is roughly constant for $E$ below and is zero for $E$ beyond the peak, while under the peak $f(E)$ is expected to be a continuously decreasing function of $E$. Hence $f(E)$ must have an inflection point under the peak. A simple function fulfilling this requirement is

$$f(E) = \frac{1}{2} f_0 \left( 1 - \frac{E - E_i}{w + |E - E_i|} \right), \qquad (2)$$

which is added to the well-known fit function for the undisturbed conversion lines near 1 MeV, which are based on Eq. (1). With the three parameters amplitude $f_0$, inflection point $E_i$, and width $w$, Eq. (2) covers practically all shapes that the residual spectrum under the peak can reasonably assume, from an almost linear decrease for large $w$ to a step function for $w = 0$.

A fit of Eq. (1) plus (2) with the six parameters $n_0$, $E_0$, $\sigma$, $f_0$, $E_i$, $w$ to the data gives for Eq. (1) the black, for Eq. (2) the blue curve in Fig. 4b. For the latter, the inflection point $E_i$ lies near the peak center $E_0$ (within a few percent), while the width $w$ is comparable to the peak's FWHM. This means that beyond $E_0$, the backscatter fraction approximately follows the downward shape of the conversion peak and is thus roughly a constant fraction of its downward shape. With $\chi_0^2 = 3.8$ for 38 degrees of freedom, the fit is not perfect, but gives a reliable value $R = 281(7)$ p.e./MeV for the yield. Further sources of systematic error will increase this 2.5% fitting error to up to 9%, as will be discussed in Section 7.

In the 6-dimensional parameter space there are two other local $\chi^2$ minima of similar depth. The three minima are well separated from each other by saddle points that are 20 to 30 units of $\chi^2$ above the minima. Both local side minima can be excluded on physics grounds: In one, called the lower minimum, the backscatter fraction $f(E)$ has its inflection point $E_i$ just where the peak begins to rise, such that almost no backscattered intensity is under the peak. The other, called the upper minimum, has $E_i$ at the upper end of the peak's spectrum where its intensity goes to zero again, such that the backscatter fraction is practically constant under the peak, and increases to 100 percent of the intensity at the peak's end. Both cases are equally unlikely.

The yield at the false $\chi^2$ minima is $R = 253(5)$ p.e./MeV for the lower, and $R = 348(8)$ p.e./MeV for the upper minimum, 10% and 24% off the true $R = 281(7)$ p.e./MeV, which shows the importance of the backscatter correction. Note that, if Eq. (2) is replaced by the similar-looking function $f(E) = \frac{1}{2}f_0\{1 - 2\arctan[(E - E_i)/w]/\pi\}$, the result is practically the same, with the three $\chi^2$ minima found at the same positions.

## 6. Measurement of photon losses

In order to reduce the number of unknown parameters, the photon transmission parameters $R_t$, $R_f$, and $\lambda$ introduced in Section 2 were measured in separate experiments. Reflectivity $R_t$ under total reflection within the scintillator was measured with a setup sketched in **Fig. 5a**, apparently a standard method, see Ref. [31]. The beam of a green laser pointer was totally reflected repeatedly between the two inner faces of a 60×60×5 mm³ scintillator, with no reflector foils. The number of reflections was varied via the angle of incidence $\theta$. The transmission of the light was measured with a 10×10 mm² photodiode (FDS1010, Thorlabs) coupled directly to a nano-ammeter.

**Fig. 5b** shows the result for the total-reflection measurement. The scatter of the red data points reflects the limited stability of the ammeter. The black transmission curve is calculated using Fresnel theory. For each additional reflection, a step appears in the transmission curve. Data points above $\theta = 80°$ suffer from incomplete illumination of the scintillator face and are excluded from the fit. The fit to the data gave a reflectivity $R_t = 96.6(3)\%$ per total reflection, and an absorption length of $\lambda = 1.3(4)$ m, with errors as given by the fit routine.

In a similar setup, foil reflectivity $R_f$ was measured (not shown) via multiple reflections from two parallel, 8 cm long Vikuti reflector foils, with no scintillator in between. Only at the end section of the setup, a 1.5 cm short piece of light guide was inserted (and included in the calculation) to avoid the bright light reemitted from the scissor-cut end of the reflector foil. The result is $R_f = 95.2(3)\%$ per reflection.

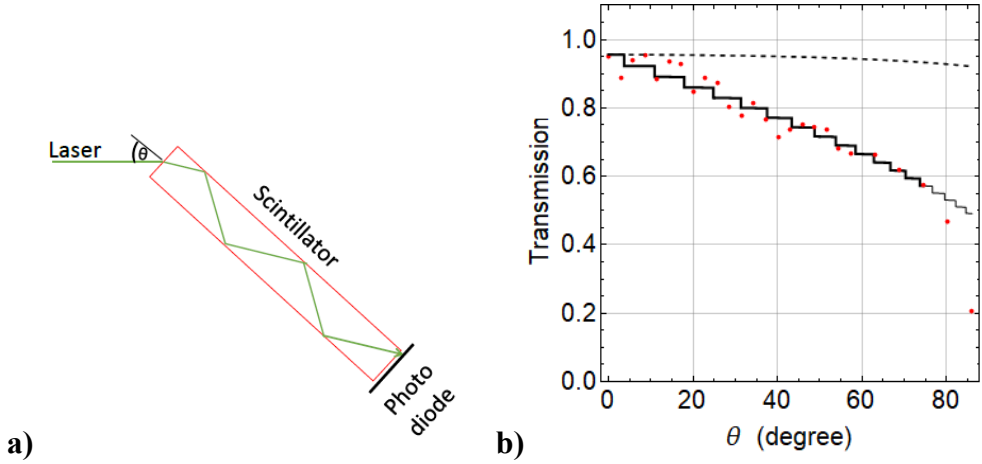

**Fig. 5.** Measurement of photon losses upon total reflection. **a)** Sketch of the setup, with totally reflecting surfaces in red. **b)** Calculated photon transmission (black) fitted to the measured data points (red) as a function of the angle of incidence $\theta$. The thick curve gives the fit interval. The dashed curve indicates the absorption losses in the scintillator obtained from the fit, starting at 95% transmission for normal incidence $\theta = 0°$.

The laser pointer had a wavelength of 515 nm, while the peak sensitivity of the scintillator is at 423 nm. From the measurements in [32] we know that the Vikuti foil reflectivity $R_f$ is wavelength independent for this range, but this is not the case for total reflectivity $R_t$ and absorption length $\lambda$ of the scintillator. From the loss data in [31], taken at three different wavelengths, we deduce that the absorption length at 423 nm is diminished to about one half or $\lambda \approx 0.7$ m, while total reflection losses $1 - R_t$ increase by about 45%, diminishing reflectivity from our measured $R_t = 96.6\%$ to about $R_t = 95\%$.

In future experiments, $R_t$ can certainly be improved, because our scintillators showed visible signs of use. For both reflectivity measurements, our assumption of a negligible angular dependence of $R_t$ and $R_f$ was justified a posteriori by the agreement of theory and experiment.

## 7. Optimum number of SiPMs

Both photoelectron yield $R$ and rate $r_{dark}$ of dark counts increase with the number $N$ of SiPMs coupled to a scintillator. For the neutron decay experiments with PERC, scintillators of typical size 120×120 mm² are planned [33]. Scintillators of this size require a large number $N$ of SiPMs in order to detect most photons before they are lost. The question then is whether, with a specific dark rate of up to 100 kHz/mm², both a high yield $R$ and a low detection threshold can be realized simultaneously. To find out, we measured $R$ as a function of $N$ and compared the result to expectations.

Measurements of $R(N)$ were done for two different scintillator sizes, one of 20×20 mm² area, the other of a four times larger area 40×40 mm², both of thickness 5 mm, with $N \leq 5$ SiPMs for which dark noise can still be handled. This allows extrapolation to the larger scintillators and larger $N$ needed in neutron decay spectrometers, for which dark noise cannot be handled yet, possible solutions being discussed in the next section.

The yields $R$ were obtained from the widths of the 1 MeV $^{207}$Bi conversion line measured at overvoltage +4.5 V as described in Sections 4 and 5. An initial light yield of $Y = 10^4$ photons per MeV deposited electron energy and a SiPM photon detection efficiency of $PDE = 40\%$ were taken from the data sheets of the suppliers. The measured photon loss parameters $R_f$, $R_t$, and $\lambda$ from the previous section were used. **Fig. 6** shows the measured and calculated yields $R$ for different numbers $N$ of SiPMs.

With these parameters given, the only free parameter was an overall device efficiency $\varepsilon$. The efficiency needed for agreement with the data was $\varepsilon = 0.51$ for the 20 mm and $\varepsilon = 0.69$ for the 40 mm wide scintillator. They differ probably due to different histories of use and wear. The unaccounted losses $1 - \varepsilon$ are in part also due to the refraction mismatch of scintillator, optical grease, and light sensor, which can lead to significant losses, see Fig. 2.21 in Ref. [25].

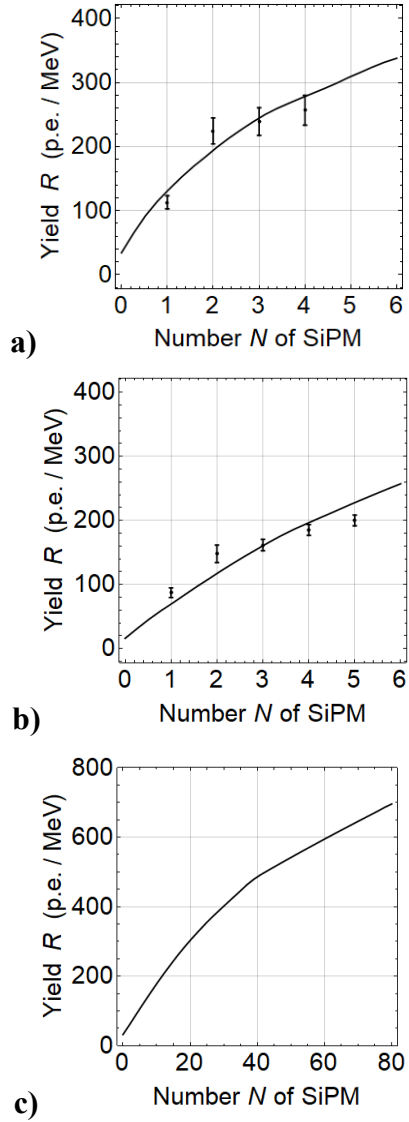

**Fig. 6.** Yields *R* of scintillators of different sizes as functions of the number *N* of SiPMs installed, **a)** for a 20×20×5 mm³, **b)** for a 40×40×5 mm³, **c)** for a 120×120×12 mm³ plastic scintillator, note change of scale. For details see text.

The errors in Fig. 6 were derived as follows. For a given scintillator-plus-SiPM combination, repeated measurements of yield *R* with no other changes gave reproducible results within ±5% (one sigma) statistical variation. A ±6.5% variation of *R* is observed for different SiPM chips, which probably is not a real effect, because the same variation is seen upon a mere reinstallation of one and the same SiPM. The typical errors of the signal fitting procedure were smaller than this by about a factor of two. From these results an overall error of ±9% was estimated for the data in Fig. 6a,b. The above variations were found in a long series of test measurements. The data in Fig. 6, on the other hand, were taken all in one final run under identical conditions. The agreement between data and theory, though not perfect, is sufficient for our following rough estimate.

Fig. 6c shows $R(N)$ calculated with the same parameters as in Fig. 6b (i.e., $\varepsilon = 0.69$), but for a scintillator of size 120×120×12 mm$^3$. In this panel, $N$ is going up to the maximum number of SiPMs that can be accommodated on two opposing narrow faces of this scintillator, namely, $N_{max} = 80$ if we take SiPM chips of 6×6 mm$^2$ area that are now available. The slight kink in the curve for $N = N_{max}/2$ is due to shadowing of virtual SiPMs by the real SiPMs.

Yield $R$ was reproducibly found to be independent, within ±6%, of overvoltage in its range from +2.5 V to +5.5 V. On the other hand, in this range, the data sheet of our SiPM shows an increase of photon detection efficiency, from 31% at +2.5 V to a near-saturation value of 43% at +5.5 V, both at 420 nm wavelength. Apparently, for our diodes saturation already sets in at lower overvoltage. As the curves $R(N)$ were all taken at a constant overvoltage of +4.5 V, this unexplained feature does not affect the conclusions drawn from Fig. 6.

If not two but all four narrow faces of this large scintillator are covered with SiPMs, no reflector foils are needed, and photon losses become negligible because the number of total reflections on the large inner faces is limited to five for the given scintillator dimensions. The expected yield then is near the limiting value $R = \varepsilon\, q\, (\Omega/4\pi)\, Y = 0.69 \times 0.35 \times 0.77 \times 10^4 = 1860$ p.e./MeV. For a virgin scintillator with a better overall efficiency $\varepsilon$, this value can probably be increased to above 2000 p.e./MeV. To compare: In present neutron decay experiments, values of up to 500 p.e./MeV are reached when scintillators are read out on all four sides with conventional PMTs, limited by additional losses in the light guides and the dynodes of the PMT.

## 8. Suppression of SiPM dark counts

In our above example of $N = 160$ SiPMs attached all around the narrow faces of a 120×120×12 mm$^2$ scintillator plate, the rates of multiple random coincidences of dark counts will be so high that they can no longer be blocked by an acceptable discriminator threshold. But how can up to $N = 20\,000$ SiPMs be operated in parallel in the high energy experiments, as quoted at the end of Section 2? This is possible because signals from a beam of particles are counted in multiple coincidence with signals coming from other detectors upstream, and are required to fall into a time window $\Delta t < 100$ ps, short enough that hardly any random dark events intervene.

If the total rate of uncorrelated events is $r_{tot}$, the rate $z_n$ of random $n$-fold coincidences is given [34, 35] by the ratio

$$\frac{z_n}{r_{\text{tot}}} \approx n(\Delta t\, r_{\text{tot}})^{n-1}. \tag{3}$$

Hence, for sufficiently small $\Delta t\, r_{\text{tot}}$ and sufficiently large $n$, the random rate $z_n$ is strongly suppressed. With each additional coincidence requirement, $z_n$ drops in steps of width $E_{1\text{pe}} = R^{-1}$ and relative intensities

$$\frac{z_{n+1}}{z_n} \approx \frac{n+1}{n}\Delta t\, r_{\text{tot}} \approx \Delta t\, r_{\text{tot}}. \tag{4}$$

The remaining random coincidences can then be suppressed by setting a discriminator threshold at $nE_{1\text{pe}}$.

In neutron decay experiments, no such preceding signals exist for coincidence suppression of dark counts. A solution to this problem is that multiple coincidences can be required not only for signals from different separate scintillators, as done in the high-energy experiments, but also for signals coming from several light detectors coupled to one and the same scintillator. Indeed, most neutron decay experiments listed in the Introduction use several PMTs for the readout of a single scintillator and require that the signals of at least two of them are in coincidence, in order to suppress the small PMT dark rate of typically several kHz.

To generalize this method to arbitrary numbers $n$ of coincidences, let a number $N$ of SiPMs of area $a$ each be installed on a single scintillator, with $N$ being a multiple of $n$. With a specific dark rate $r_0$ per detector area, the total dark rate is $r_{\text{dark}} = Nar_0$. Let the $N$ detectors be subdivided into $n$ groups with $m = N/n$ SiPMs per group, and require that the signals are in $n$-fold coincidence within a time window $\Delta t$. For sufficiently small $\Delta t$, the fraction of random coincidences from Eq. (3) is again suppressed by a detector threshold of $nE_{1\text{pe}}$.

The total number of photoelectrons $p$ registered in a SiPM is Poisson-distributed as $P_{\bar{p}}(p) = \bar{p}^p \exp(-\bar{p})/p!$ about the mean value $\bar{p} = \tilde{E}R$, with yield $R$ in p.e./MeV and $\tilde{E}$ in units of MeV. The probability that a signal is lost due to a statistical fluctuation to $p = 0$ counts in one of the $n$ groups then is $nP_{\bar{p}/n}(0) = n\exp(-\bar{p}/n)$ and can also be made very small.

To give a numerical example: The dark rate of commercial SiPMs has diminished over the years and is now at about $r_0 = 40$ kHz/ mm$^2$. A large scintillator, fully equipped on all sides with a total number of $N = 160$ SiPMs of area $a = 6\times 6$ mm$^2$ each has a dark rate $r_{\text{dark}} = Nar_0 = 2.3 \times 10^8$ s$^{-1}$. A time window of $\Delta t = 0.5$ ns gives $\Delta t\, r_{\text{dark}} = 0.115$. Divide the SiPMs on the scintillator into $n = 10$ groups of $N/n = 16$ SiPMs each. The fraction of random coincidences from Eq. (3) then is $z_{10}/r_{\text{dark}} = 3.6 \times 10^{-8}$, or $z_{10} = 8$ s$^{-1}$, which decreases further, if necessary, upon lowering $\Delta t$ or increasing $n$. For a lowest threshold energy $E = 30$ keV and $R = 2000$ p. e./MeV, we have $\bar{p}/n = 6$, and the loss fraction $nP_{\bar{p}/n}(0) = 10 \times \exp(-6) = 0.025$ is small and reliably calculable. This threshold is quite reasonable, because at present in neutron decay evaluations spectral fits begin well above a $\beta$ energy of $E = 100$ keV.

Hence, the dark current problem in neutron decay can be solved in a similar way as in the high-energy experiments. Furthermore, in contrast to the high-energy case, neutron decay experiments have extremely low radiation background and do not have the problem of radiation aging of their detectors.

However, there remains a problem: Eqs. (3) and (4) hold only for uncorrelated dark counts, and there remains the problem of dark counts that suffer optical cross talk or are correlated with afterpulses, which the datasheet of our SiPMs give as several percent of all events. The correlated dark counts may push the required detector threshold well beyond the value $nE_{1pe} = nR^{-1}$ needed to suppress uncorrelated random $n$-fold coincidences. Correlated noise is best taken account of by introducing an excessive noise factor (ENF), which can well surpass the ENF due to uncorrelated noise, as is discussed in Ref. [17].

To estimate the effect of correlated noise on the required detection threshold in $\beta$ spectroscopy, instead of entering the theory of correlated noise, we take a more practical approach and have a look on the results of experimental γ spectroscopy with SiPM readout. Large NaI scintillators of size up to 2" by 2" have been read out by arrays of up to 64 SiPMs of area 6×6 mm$^2$ each, and were found to be free of dark noise above a γ energy of 15 keV [21], [22], which is highly sufficient for our requirements. Therefore, one can be optimistic that the problem of correlated noise can also be handled in neutron decay studies with SiPMs. Furthermore, the field is moving rapidly, in the words of Ref. [17]: "SiPM developments are in fast progress in many directions to meet advanced demands of applications", such as medical, high energy and optical applications.

For neutron decay experiments, as well as for $\beta$ decay spectroscopy in general, it will be interesting to have a scintillator plate simply equipped on its narrow faces with rows of small SiPM chips, with the potential of a 4-fold photoelectron yield, or doubled energy resolution, as compared to the best PMT figures.

# 9. A complete experiment on electron backscattering.

The use of SiPMs will also facilitate auxiliary experiments on the characterization of electron scintillators used in neutron decay. Earlier such studies were on the point spread function of the electrons after magnetic transport [36], [37], and on the response function of electron detectors with a method using electron time-of-flight [30]. We want to continue with an experiment on electron backscattering, another dominating sources of error.

Electron backscattering from plastic scintillators had been studied before, though limited to normal incidence [38] or to an integrated angular range [39] of the incoming electrons. Our planned experiment will be complete in the sense that it covers all relevant electron variables in a single run, including low electron energies as well as grazing angles of incoming and outgoing electrons, where Monte Carlo simulations [40], [41] have not yet been sufficiently controlled experimentally and where different codes give different predictions, see Fig. 1a in Ref. [38].

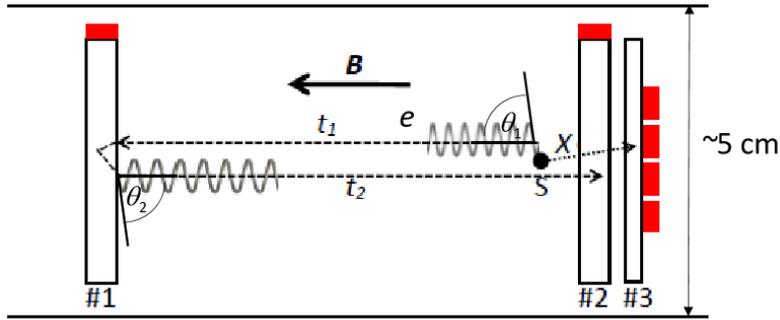

**Fig. 7.** Arrangement of SiPMs (red) in a complete experiment on electron backscattering. The electrons, emitted by the source S under angle $\theta_1$, gyrate along the uniform magnetic field $B$ until they reach detector #1 under the same angle $\theta_1$, after a measureable time-of-flight $t_1$. The electrons that are backscattered under angle $\theta_2$ reach detector #2 after a ToF $t_2$. The deposited electron energy is measured in both detectors. Time zero is given by an x or γ ray detected in a LaBr$_3$ scintillator #3. For details see text.

In the planned backscatter experiment, only small scintillators and with it small numbers $N$ of SiPM are required, such that dark counts can be suppressed in a simple coincidence setup. As shown in **Fig. 7**, an electron is backscattered from a scintillator #1 and is registered in a scintillator #2, both of ~40 mm diameter and ~5 mm thickness, and a distance $z_0 \sim 10$ cm apart from each other. The measured energies deposited in the scintillators sum up to $E = E_1 + E_2$. In the uniform guide field $B$, the angle of emission $\theta_1$ from source S of is the same as the angle of incidence on #1. With the measured ToF $t_1$, this angle of incidence is calculated to $\theta_1 = \arccos[z_0/(t_1 v_1(E))]$, using standard formulae $v_1 = c\beta$, $\beta = cp/W$, $W = E + mc^2$, $cp = \sqrt{W^2 - (mc^2)^2}$, and similarly for $\theta_2 = \arccos[z_0/(t_2 v_2(E_2))]$.

The ToF start pulse is provided by a prompt photon from a $\beta$-$\gamma$ emitter, or from a conversion electron plus x ray emitter. In the case of a $^{207}$Bi (30 y) source, conversion electrons are accompanied by x rays from the K shell of mean energy 77 keV, which are registered in the LaBr$_3$ crystal #3. The electrons detected in #2 do not reach #3, while the x rays cross #2 with almost no losses. The ratio of triple coincidences (#1,#2,#3) to double coincidences (#1,#3) will give the backscattering coefficient as a function of the parameters $E, E_2, \theta_1, \theta_2$, all in one single run. In the $\beta$-$\gamma$ case, electron energy it is measured as $E = E_1 + E_2$, while for an electron conversion line $E$ is known and hence overdetermined.

With the detector arrangement of Fig. 7, with 20 ns coincidence time and a backscattering coefficient of 10%, a more detailed study shows that for every 1000 conversion electrons at 1.0 MeV, 12 true and 0.2 random triple coincidences (#1,#2,#3) and 120 true and 0.14 random double coincidences (#1,#3) will occur. Gamma ray and electron backgrounds are negligible. In particular, triggering with an x ray guarantees that there is no coincident background from the alternative Auger electron transition. The angular sensitivity obtained with this method was discussed in Section 4 of [42]. The experiment can be done in the long uniform field region with $B$ = 2 T of the upcoming PERC instrument.

A feasibility test for x ray detection was made with an encapsulated 20×20×3 mm$^3$ LaBr$_3$(Ce) crystal of density 5.1 g/cm$^3$ (Ost Photonics Co.). The hygroscopic crystal had a 1 mm Al window and was read out with a conventional setup of light guide plus PMT. The 77 keV x-ray K-line had 34 keV FWHM, corresponding to $R$ = 360 p.e./MeV, and was almost background-free. The crystal is hence well suited to give the start pulse in the backscatter project.

Further improvement in energy resolution in $\beta$ spectroscopy can possibly be obtained with LYSO crystals as $\beta$ scintillation detectors. A $^{207}$Bi electron spectrum was taken with a bare 20×20×1 mm$^3$ LYSO (Lutetium-yttrium oxyorthosilicate) crystal of density 7.4 g/cm$^3$ (Shalomeo Co.). Its photon yield is three times that of plastic scintillators, and scintillator plates are available in sizes of up to 100×200 mm$^2$ [43]. Their large refractive index $n$ = 1.83 increases the usable solid angle of photon emission to $\Omega/4\pi = 0.88$. There exist methods for coping with the resulting refraction mismatch, see references in [44].

However, with lutetium's high element number $Z = 71$, the price to pay is a higher sensitivity for γ background and a larger backscatter fraction.

## 10. Summary

The readout of plastic scintillators with silicon photomultipliers (SiPMs) was studied for use in neutron decay experiments. A simple tool based on virtual images was developed to model the combined scintillator-SiPM system, see Fig. 1. A method was found to extract the photoelectron yield from conversion electron signals in the presence of electron backscattering, see Fig. 4. For the scintillators investigated, photon reflection and absorption losses were measured separately, with results shown in Fig. 5. The photoelectron yield expected in a neutron decay experiment with SiPM readout, not yet measurable due to extreme rates of dark counts, was estimated from the photon yield measured and calculated as a function of the number of installed SiPMs, given in Fig. 6. Finally, a method for suppressing dark counts in SiPMs was proposed, which will make the SiPM-scintillator system practicable for neutron decay and other $β$ decay experiments. We shall first use such a system in a ToF-based complete experiment on electron backscattering, sketched in Fig. 7.

## Acknowledgements

I thank S. Bachmann and U. Schmidt from Heidelberg University and B. Märkisch from the Technical University Munich for helpful discussions.